\documentclass{emulateapj}
\usepackage{apjfonts}
\usepackage{amsmath}

\def\mr#1{\mathrm{#1}}

\newcounter{ichi}
\newcounter{ni}
\newcounter{san}
\slugcomment{Accepted for publication in ApJ}

\shorttitle{Effects of CIB on Delayed Gamma-Rays from GRBs}
\shortauthors{Murase, Asano, and Nagataki}

\begin{document}


\title{Effects of Cosmic Infrared Background on\\
High-Energy Delayed Gamma-Rays from Gamma-Ray Bursts}


\author{Kohta Murase\altaffilmark{1}, Katsuaki Asano\altaffilmark{2},
Shigehiro Nagataki\altaffilmark{1,3}}


\altaffiltext{1}{YITP, Kyoto University, Kyoto, Oiwake-cho,
Kitashirakawa, Sakyo-ku, Kyoto, 606-8502, Japan}
\altaffiltext{2}{Division of Theoretical Astronomy, NAOJ, Osawa
2-21-1, Mitaka, Tokyo, 181-8588, Japan}
\altaffiltext{3}{KIPAC. Stanford University, P.O. Box 20450, MS 29, 
Stanford, CA, 94309, USA}


\begin{abstract}
Regenerated high-energy emissions from gamma-ray bursts (GRBs) are studied 
in detail. If the intrinsic primary spectrum extends to the TeV range,
these very high-energy photons are absorbed by the cosmic infrared 
background (CIB). Created high-energy electron-positron pairs
up-scatter mainly cosmic microwave background (CMB) photons, and 
secondary photons are generated in the GeV-TeV range. These secondary 
delayed photons may be observed in the near
future, which are useful for a consistency check of the intrinsic
primary spectrum. In this paper, we focus on effects of the
CIB on delayed secondary emissions. In particular, we show that not only 
up-scattered CMB photons but also up-scattered CIB ones are important,
especially for low redshift bursts. They also give us additional information 
on the CIB, whose photon density is not definitely determined so far.
\end{abstract}





\section{Introduction}
Gamma-ray bursts (GRBs) are highly relativistic astrophysical phenomena
located at a cosmological distance. Observed gamma-ray spectra are
typically highly nonthermal and extended to the MeV range or
above. The relativistic shock scenario is one of the leading scenarios
to reproduce such spectra (see reviews, e.g., \cite{Mes3,Zha1}). The 
internal shock model is one of widely accepted
models. In this model, the GRB prompt emission is explained by
electromagnetic radiation from 
relativistic electrons accelerated in shocks generated by collisions 
among the subshells. Theoretically, several emission
mechanisms of GeV-TeV emission have been proposed. The synchrotron
self-inverse Compton mechanism (SSC) is one of them
\cite{Pap1,Dai2,Gue1,Pee1,Cas1}. 
While such models belong to leptonic scenarios, hadronic scenarios are
also possible. In GRBs, protons can be accelerated up to the ultra high 
energy region. high-energy protons can not only radiate synchrotron
photons \cite{Tot1} but also produce electron-positron pairs, pions and muons 
via the photohadronic process \cite{Wax1,Asa1,Asa4}. Synchrotron radiation by
electrons, positrons and muons 
can also contribute to resulting spectra \cite{Der2,Asa2}.
Neutrino detection would be strong evidence
of baryon acceleration and expected by future
neutrino detectors such as IceCube \citep{Wax1,gue01,gue04,der03,KM1,Asa3}.

Sufficiently high-energy photons, including photons originating from protons,
make pairs mainly via $\gamma \gamma \rightarrow e^{+} e^{-}$
interaction in the subshells and cannot escape from the source. 
As a result, the intrinsic high-energy cutoff of
GRB spectra is usually determined by the optical depth for pair
creation, which largely depends on bulk Lorentz factors of
subshells \cite{Lit1,Raz1}. Even if such high-energy photons escape 
from the subshells, these photons may suffer from interactions with 
the cosmic infrared background (CIB) and be largely absorbed
especially for high-$z$ GRBs such as $z \gtrsim 1$. 
Hence, the detection
of TeV photons will be very difficult, unless the GRB location is
nearby. The secondary electron-positron pairs generated by
attenuation are very energetic, so that they
up-scatter numerous cosmic microwave background (CMB)
photons by the inverse-Compton (IC) process. 
Such secondary photons will be observed as delayed GeV emissions
\cite{Che1,Dai1,Dai2,Gue1,Wan2,Raz1,And1,Fan1,Cas1}. 
The delayed secondary emission is indirect evidence of the 
intrinsic TeV emission as well as a clue to probing
the intergalactic magnetic (IGM) field \cite{Pla1}.  
Such delayed secondary emission has been discussed in terms of not only the
internal shock model but also the external shock model
\cite{Mes1,Mes2,Der3,Zha2,Der4,Wan1,Wan2,And1}. They could be 
distinguishable by the distinct spectral evolution behavior. 

Photon attenuation due to the CIB is very useful as an indirect
probe of the CIB, whose photon density is not satisfactorily determined. 
The direct observation of the CIB is difficult, because of the bright 
foreground emission associated with zodiacal light as well as emission
from our Galaxy.
COBE DIRBE and COBE FIRAS have succeeded in highly significant detections 
of the residual diffuse infrared background, providing an upper bound on the 
CIB in the infrared regime. On the other hand, galaxy counts have 
given a lower bound on the CIB at wavelengths, where no COBE data are 
available. Despite the dramatic progress in observations achieved by 
Infrared Astronomical Satellite (IRAS), Infrared Space 
Observatory (ISO) and the Submillimeter Common-User Bolometric
Array (SCUBA), the mid-infrared (MIR) and far-infrared (FIR)
observations do not reach a level of optical and near-infrared (NIR)
bands, which can be explained by direct stellar emission. In this
sense, we have not determined the spectral energy distribution (SED)
of the CIB with sufficient accuracy yet. For recent reviews, see 
Hauser \& Dwek (2001) and Kashlinsky (2006). Stecker et al. (1992)
proposed that one can use photon attenuation in blazars to
determine the intensity of the CIB, if we know intrinsic spectra of
blazars. Subsequent studies used observations of TeV emissions from blazars
(for one of the latest examples, Aharonian et
al. (2006)). Conversely, we could obtain information on the intrinsic
primary spectrum such as the intrinsic high-energy cutoff, if we know the CIB 
accurately. Similarly to cases of blazars, we can expect to make use of GRBs as
a probe of the CIB. However, owing to uncertainties in GRB intrinsic spectra,
the depletion due to the CIB in high-energy spectra is
hard to be estimated.

Observationally, GeV photons have been detected from some GRBs
with the EGRET detector on the Compton Gamma-Ray
Observatory. Especially, EGRET detected prolonged GeV emission from GRB
940217 \cite{Hur1} and GRB 930131 \cite{Som1}. 
Although we do not know the highest energy in GRB spectra
observationally, theoretical consideration and simple extrapolation of 
GRB spectra enable us to expect TeV photons from some GRBs.
Furthermore, the tentative detection of an excess of TeV
photons from GRB 970417a at the 3 $\sigma$ level has been claimed with
a chance probability  $\sim 1.5 \times {10}^{-3}$ by the water
\v{C}herenkov detector Milagrito \cite{Atk1}, although Milagro has not
observed such signals so far \cite{Mil1}.
Another possible TeV detection of
GRB 971110 has been reported with the GRAND array at the $2.7 \sigma$
level \cite{Poi1}. Staking of data from the TIBET array for a large
number of GRB time window has led to an estimate of a $\sim 7 \sigma$ composite
detection significance \cite{Ame1}. 
Further observations of such very high-energy gamma-ray signals by 
MILAGRO, MAGIC \cite{Mir1}, VERITAS \cite{Hol1}, HESS \cite{Hin1} and 
CANGAROO\Roman{san} \cite{Eno1} might enable us to detect the signals
in the near future. However, the photon detection in the TeV range can
be expected only for nearby events, since high-energy gamma-rays will
suffer from attenuation by the CIB. 
On the other hand, future detectors such as Gamma-Ray Large Area Space 
Telescope (GLAST) will detect many GRBs around the GeV range. They
will also enable us to discuss not only prompt primary emissions but also
delayed secondary emissions produced via IC up-scattering.

It depends on the strength of IGM field whether we can observe such
delayed gamma-ray signals or not. For electron-positron pairs produced
by attenuation of 1 TeV primary photons, the sufficiently 
strong IGM field $B \gtrsim {10}^{-16}$ G leads to 
the large magnetic deflection of pairs, $\theta _B \gtrsim 1$, and long time
delay of secondary photons, $\Delta t_B \gtrsim {10}^{3}$ s, so that 
the secondary gamma-ray flux will be suppressed. On the other hand,
the weak IGM field enables us to have possibilities to detect 
delayed secondary gamma-ray signals. 
In this paper, we study these two extreme cases, the weak IGM field case
and strong IGM field case. In
the former case, the magnetic deflection time is not so important and other
time scales such as the angular spreading time scale are more
important for relevant energies. In the latter case, it becomes difficult
to observe delayed secondary gamma-rays directly from each burst, but 
they contribute to the diffuse gamma-ray background.

In this paper, we study delayed secondary gamma-ray spectra from GRBs
most quantitatively by numerical simulations as
well as an approximate formula.
We focus on effects of the CIB by including the contribution from up-scattered 
CIB photons (hereafter, USIB photons)
to delayed secondary spectra, which can extend delayed spectra to
higher energies but has not been 
emphasized in previous studies. Such a study
would be important for GRBs that can emit $\sim$ TeV emissions
in order to know the intrinsic feature of the source. In addition,
it would be useful to obtain information on the CIB more
quantitatively. Not only GLAST but also MAGIC and VERITAS
might detect up-scattered CIB photons in the near future.

This paper is structured as follows. In \S \ref{subsec:a}, we
explain the models of intrinsic GRB spectra and we
describe the delayed emission mechanism in \S \ref{subsec:b}.
The CIB model we use in this paper is explained in \S 2.3.
We show the method to estimate the diffuse gamma-ray background
due to GRBs in \S 2.4.
In
\S 3, we show the results. Finally, our summary and discussion
are described in \S 4.

\section{\label{sec:1}THE MODEL}
\subsection{\label{subsec:a}Model of Intrinsic Spectra}
Throughout the paper, we focus on long GRBs with the typical duration
$T \sim 10-100$ s. Widths of individual pulses vary in a wide range. 
Typical pulses have the duration with $\delta t \sim 0.1-10$ s and shortest
spikes have millisecond or even sub-millisecond widths. The internal
shock model, in which gamma-rays arise by the internal dissipation of 
relativistic jets, can reproduce such wide range variability.  
However, the simple synchrotron model cannot explain
several properties of prompt emission (see, e.g., M\'esz\'aros 2006). 
In this paper, we do not consider these open problems on the prompt
emission mechanism.
The observed photon spectrum
is well approximated by a broken power-law, 
$dN_{\gamma}/dE_{\gamma} \propto{(E_{\gamma}/
{E}_{\gamma}^{\mr{b}})}^{-\alpha}$ for $E_{\gamma}^{\mr{sa}} 
< E_{\gamma} < E_{\gamma}^{\mr{b}}$ and $dN_{\gamma}/dE_{\gamma} \propto
{(E_{\gamma}/{E_{\gamma}}^{\mr{b}})}^{-\beta}$ for $E_{\gamma}^\mr{b} 
< E_{\gamma} < E_{\gamma}^{\mr{max}}$, 
where $E_{\gamma}^{\mr{sa}}$ is the synchrotron
self-absorption cutoff,
and $E_{\gamma}^{\mr{max}}$ is the intrinsic high-energy cutoff.

The intrinsic high-energy cutoff $E_{\gamma}^{\mr{max}}$ is 
typically determined by the opacity of two-photon annihilation 
into an electron-positron pair.
In the internal shock scenario, it is easy to see that TeV photons
can escape from the subshells if an internal collision radius and/or 
a bulk Lorentz factor are large enough \cite{Lit1}.

In sufficiently high-energy ranges, gamma-rays due to
electron SSC, proton synchrotron and
charged-meson/muon synchrotron
can contribute to intrinsic spectra via cascade processes. 
The resulting spectra are complicated and
the study on them is beyond scope of this paper. We will
investigate not the intrinsic emission but the delayed emission in
detail. The delayed emission depends 
on the amount of attenuated photons and would
not be so sensitive to the detail of the shape of intrinsic spectra
with a given $E_{\gamma}^{\mr{max}}$.
Throughout the paper we adopt three models
with total isotropic energy $E_{\mr{iso}}={10}^{53}$ ergs for
calculation. Model A: a broken-power law spectrum with $\alpha=1$ and
$\beta=2.2$,
$E_{\gamma}^{\mr{b}}= 300$
keV and $E_{\gamma}^{\mr{max}}=1$ TeV. Model B: the same as model A
but $E_{\gamma}^{\mr{max}}=10$
TeV. Model C: a numerically calculated spectrum obtained by Asano \&
Inoue (2007).
They perform
Monte Carlo simulations including synchrotron radiation, Compton
scattering, pair creation, synchrotron self-absorption and
particles originating from protons
such as electrons, positrons, muons and pions.
We adopt one of their numerical results as model C. For details, see Asano
\& Inoue (2007). 
Parameters adopted to obtain the spectrum in model C are
energy per subshell $E_{\mr{sh}}={10}^{50}$ ergs, 
$E_{\gamma}^{\mr{b}}= 300$ keV, an internal collision radius
$r={10}^{15}$ cm and a Lorentz factor $\Gamma=100$.
The magnetic energy density $U_{B}$ is assumed to be 
$0.1 U_{\gamma}$, where $U_{\gamma}$ is the photon energy density in the subshell. 
In Fig. 1, we show the intrinsic spectra for the three models we adopt.
The second peak of model C in Fig. 1 is due to SSC.
Above this peak energy, photon absorption due to pair creation is crucial.
The intrinsic GRB duration (defined in the local rest frame) is set to 
$T^{\prime}=50$ s.
\begin{figure}[tb]
\includegraphics[width=\linewidth]{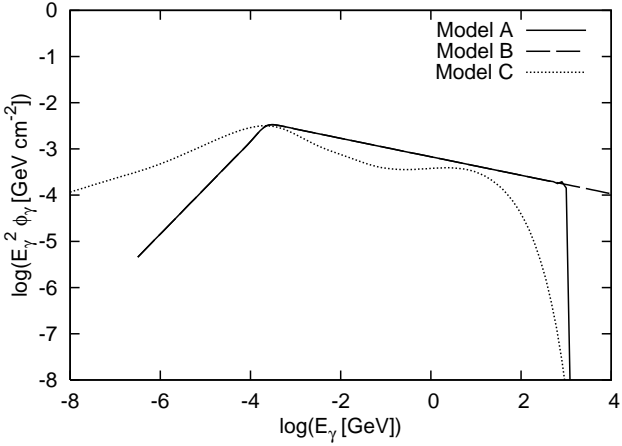}
\caption{\label{Fig1} The intrinsic primary spectra we use in this
paper for each model. The spectra are normalized by the fluences from 
a source at $z=1$.
Model parameters are described in the text.}
\end{figure}

\subsection{\label{subsec:b}Delayed Emission}
For typical
GRBs at the source redshift $z=1$, most high-energy photons above $\sim 70$ GeV
produce electron-positron pairs. The produced high-energy pairs cause
delayed high-energy photon emission via IC up-scattering of CMB and CIB
photons.
The duration of such delayed emission is determined by several effects 
\cite{Dai1,Dai2,Wan2,Raz1,And1,Fan1};
the angular spreading, IC cooling and magnetic
deflection effects. The angular spreading time is expressed as,
\begin{equation}
{\Delta t}_{\mr{ang}} \approx (1+z) \frac{\lambda _{\gamma
\gamma}}{2{{\gamma}_{e}}^{2}c} \label{ang}
\end{equation}
where ${\gamma}_{e}$ is the Lorentz factor of secondary
electrons or positrons in the local rest frame in the Robertson-Walker metric
(hereafter, the local rest frame) at each $z$,
and $\lambda _{\gamma \gamma}$ is the photon mean free path.
The IC cooling time scale is written as,
\begin{equation}
{\Delta t}_{\mr{IC}} \approx (1+z) \frac{{\hat{t}}_{\mr{IC}}}
{2{{\gamma}_{e}}^{2}} \label{ic}
\end{equation}
where $\hat{t}_{\mr{IC}}$ is the cooling time scale
in the local rest frame.
If the magnetic deflection angle is sufficiently small, the magnetic 
deflection time is,
\begin{equation}
{\Delta t}_{B} \approx (1+z) \frac{1}{2} {\hat{t}}_{\mr{IC}}
{\theta}_{B}^{2} \label{mag}
\end{equation}
where $\theta_B = c \hat{t}_{\mr{IC}/}r_L$ is the magnetic deflection
angle and $r_{\mr{L}}$ is the Larmor radius of
electrons or positrons. Note that we have implicitly assumed
$1/\gamma_e, \theta_B \ll \theta_j$ where $\theta_j$ is
an opening angle of GRB jet.
Taking into account the GRB duration $T$, the (secondary) duration time
scale is estimated by the maximum time scale, $\Delta t= \mr{max}[{\Delta
t}_{\mr{ang}}, {\Delta t}_{\mr{IC}}, {\Delta t}_{B}, T]$.
Examples of
$\Delta t$ adopting the CIB model of Kneiske et al. (2004,
see section 2.3) are shown in Fig. 2.
In cases with a weak magnetic field such as
$B \lesssim {10}^{-(18-19)}$ G, the angular spreading time 
scale is the most important.
Of course, we
should note that we treat the averaged flux over the
duration time scale.
\begin{figure}[tb]
\includegraphics[width=\linewidth]{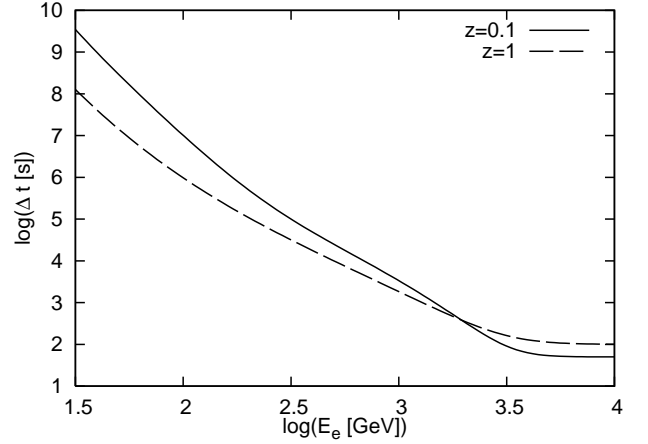}
\caption{\label{Fig2} The maximum time scales for $z=0.1$ and $z=1$ 
involved in calculating the delayed secondary spectra. $E_e$ is the electron
energy in the local rest frame. The IGM field is assumed to be
$B={10}^{-20}$ G. }
\end{figure}

The IGM field make the situation 
complicated and more careful treatments would be needed to evaluate 
the accurate time-dependent flux of delayed emission. The deleyed
emission can also be a probe of the IGM field \cite{Pla1}, but we do
not focus on this topic in this paper. We will treat the two extreme
cases. One is the weak IGM field case, where we can expect detectable 
delayed secondary gamma-ray signals from GRBs. For demonstration, we 
use an IGM field with $B={10}^{-20} \, \mr{G}$, in which ${\Delta
t}_{B}$ is not so important. The other is the strong IGM field case, 
where it becomes difficult to detect delayed signals directly 
from each GRB. We will consider this case later.
  
Let us use the local approximation that TeV gamma-rays from a GRB interact with
the CIB field at the same redshift, which is justified by numerical 
calculations later. We can obtain delayed spectra from a
burst with the source redshift $z$ by an analytic approximate formula, 
which is given by \cite{Blu1,Dai1,Wan1,Raz1,And1,Fan1},
\begin{eqnarray}
\frac{dF_{\gamma}}{dt dE_{\gamma}}(E_{\gamma}, t, z) = \int _0^t dt_{p} \int
d\varepsilon \int _{{\gamma _e}^{\mr{min}}}
^{{\gamma _e}^{\mr{max}}} 
d\gamma _e \left( \frac{dF_{e}}{dt_{p}d\gamma
_e} \right) \nonumber\\ \times  \left( 
\frac{dN_{\gamma}}{dE_{\gamma} d\varepsilon d \hat{t}_{d}} \right)
e^{-\tau_{\gamma \gamma}^{\mr{bkg}}(E_{\gamma},z)}
\hat{t}_{\mr{IC}} \frac{e^{-(t_{d}/\Delta t)}}{\Delta t}, \label{delay}
\end{eqnarray}
where the electron injection spectrum
\begin{eqnarray}
\frac{dF_{e}}{dt_{p}d\gamma _e}(E_e, t_p, z) &=& 
2 \frac{dE_{\gamma,i}}{d\gamma _e} 
\frac{dF_{\gamma,i}}{dE_{\gamma,i}dt_{p}}(E_{\gamma,i}, t_p, z) \nonumber \\ 
&\times& (1-e^{-\tau_{\gamma \gamma }^{\mr{bkg}}(E_{\gamma,i},z)}) , 
\label{dist}
\end{eqnarray}
and the photon emission spectrum per unit time due to IC scattering
\begin{eqnarray}
\frac{dN_{\gamma}}{dx d\varepsilon d \hat{t}_{d}}&=&
\frac{2 \pi r_{0}^{2} m_e c^3}{\gamma _{e}}
\frac{1}{\varepsilon} \frac{dn}{d\varepsilon} (\varepsilon,z) 
\bigl[ 2y \ln(2y) \nonumber \\
 &+& (1+2y)(1-y) + \frac{{(wy)}^2}{2(1+wy)}(1-y) \bigr], \nonumber \\
x &\equiv& \frac{E_{\gamma}(1+z)}{{\gamma}_{e} m_e c^2}, \nonumber \\
y &\equiv& \frac{x m_e c^2}{4 \varepsilon \gamma_{e}(1-x)},
\nonumber \\
w &\equiv& \frac{4 \varepsilon \gamma_{e}}{m_e c^2}.
\end{eqnarray}
Here, $dF_{\gamma,i}/dE_{\gamma,i}dt_{p}$ is the intrinsic primary gamma-ray
spectrum of GRB prompt emission, $E_{\gamma,i}=2 \gamma _e m_e
c^2/(1+z)$ is energy of primary
photons (where the source redshift is taken into account), 
$dn/d\varepsilon$ is the photon density spectrum of the CMB and CIB in
the local rest frame, and $r_0$ is the classical electron radius.
$t$ is the given observation time of the delayed emission, $t_p$ is
the time when primary photons are released, $T$ is the GRB duration,
$t_d$ is defined by $t_d=t-t_p$ and
$\tau_{\gamma \gamma}^{\mr{bkg}}(E_{\gamma},z)$ is the optical depth
against gamma-rays propagating in the universe.
The upper bound of the integration over $\gamma _{e}$ is 
determined by the high-energy cutoff of the intrinsic primary emission, 
i.e., ${\gamma _{e}}^{\mr{max}}= (1+z)E_{\gamma}^{\mr{max}}/2$. 
On the other hand, the lower bound of the integration over 
$\gamma _{e}$ is ${\gamma _{e}}^{\mr{min}}= 
\mr{max}[m_e c^2/2 \varepsilon,
{((1+z)E_{\gamma}/\varepsilon)}^{1/2}/2]$. 
We exploit Eq. (\ref{delay}) iteratively by
substituting $(dF_{\gamma}/dt dE_{\gamma}) (\exp
(\tau_{\gamma \gamma}^{\mr{bkg}})-1)$ into $dF_{\gamma,i}/dt_p
dE_{\gamma,i}$ instead of using the intrinsic primary flux. We perform such
an iterative method in order to include IC scattering by generated
pairs due to re-absorbed secondary photons.  

In the above formula, it is assumed that secondary pairs are produced
only at the source redshift and cooling time scales are evaluated with
quantities at the source redshift.
The photon emission spectrum is evaluated with the initial energy of 
secondary pairs and assumed to be constant while the pairs cool.
However, pairs are produced at various redshifts
and the cooling rate becomes lower as they cool.
Therefore, we also execute numerical
simulations including IC scattering and pair creation.
Based on the pair creation rate at each redshift
due to the CIB (partially CMB),
we follow the time evolution of the distribution functions
of primary photons $f_\gamma(E_{\gamma,i})$,
secondary pairs $f_{e}(\gamma_e)$,
and secondary photons $f_2(E_\gamma)$ from the burst time
to the present time.
The minimal time (redshift) step for
$f_\gamma(E_{\gamma,i})$ in our simulation is $dz=0.005$.
The pair cooling process is followed with a time step
${\hat{t}}_{\mr{IC}}/100$ until pairs become non-relativistic.
While $f_\gamma(E_{\gamma,i})$ decreases monotonically with time
(or remains constant for lower energy range) by attenuation,
$f_2(E_\gamma)$ does not necessarily change monotonically
especially for high-redshift sources, because of re-absorption.
IC photon spectra are calculated using the Klein-Nishina
cross section with the Monte Carlo method used in \cite{Asa1}.
Our method can precisely treat re-absorption of secondary photons
and energy loss processes of electron-positron pairs.

\subsection{\label{subsec:d}Cosmic Infrared Background}
Gamma-ray absorption due to pair creation in cosmological scales
depends on the line-of-sight integral of the evolving density of low
energy photons in the universe.
To demonstrate the effect of the CIB on delayed spectra from GRBs, we
need to exploit some model of the CIB.
The CIB should be explained by a theory from the first principles, but we
are far from this ultimate goal owing to poor knowledge about star
formation, supernova feedback, galaxy merging and so on. 
So far many models of SED of the CIB produced by stellar emission and dust
re-radiation in galaxies have been constructed \cite{Tot2,Kne1,Ste3}. 
At lower redshifts, these models agree with each other 
basically. At higher redshifts, Stecker et al. (2006) found the larger
optical depths than previously thought because of
intergalactic gamma-ray absorption motivated by the recent discovery
of active star formation taking place in young galaxies at high
redshifts. Such model uncertainties will produce corresponding 
differences.
\begin{figure}[b]
\includegraphics[width=\linewidth]{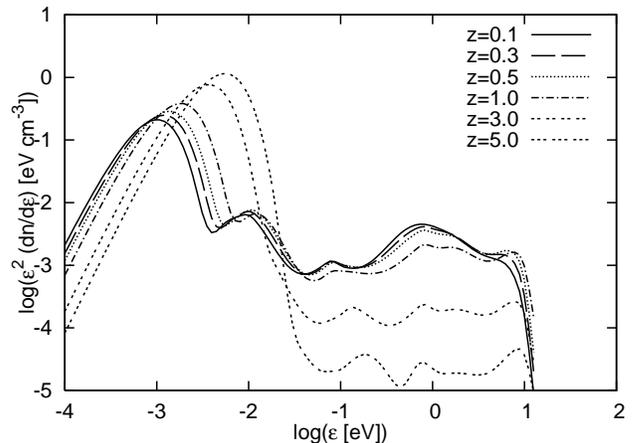}
\caption{\label{Fig3} The CMB+CIB radiation field in the comoving
frame for various redshifts, which we have used in this paper. The
data about the CIB field are taken from the best-fit model of Kneiske et al. (2004)}
\end{figure}

In this paper, we use the CIB model
developed by Kneiske et al. (2002,2004). They developed the evolving
model of the infrared-to-ultraviolet metagalactic radiation field,
based directly on observed emissivities. They specially addressed the
redshift evolution of the SEDs, which are constructed from realistic
stellar evolution tracks combined with detailed atmospheric models 
\cite{bru93}, and also taken into account effects of re-radiation 
from dusts and Polycyclic Aromatic Hydrogen molecules in the infrared.
Their model parameterizes the main
observational uncertainties, the redshift dependence of the cosmic
star formation rate and the fraction of UV radiation released from the
star forming regions.
Here, we adopt the ``best-fit model'' of Kneiske et al. (2004), which
is consistent with the data obtained from recent galaxy surveys.
In Fig. 3, we show the SED of the CMB$+$CIB we use in this paper. 
For details, see Kneiske et al. (2002,2004).
We also assume the $\Lambda$CDM cosmology with $\Omega_{m}=0.3$,
$\Omega_{\Lambda}=0.7$ and $H_0=70 \, \mr{km} \mr{s}^{-1} \mr{Mpc}^{-1}$. 
 
Given the SEDs, we can calculate the mean free path of high
energy gamma-rays for pair creation or pair creation rate at each redshift.
Especially, using the cross section of pair creation $\sigma_{\gamma \gamma}$,
the optical depth of the universe is written as,
\begin{equation}
\tau_{\gamma \gamma}^{\mr{bkg}} = \int _{0}^{z} dz  \left|
\frac{cdt}{dz} \right|
 \int d \cos \theta \frac{1-\cos \theta}{2} \int
d\varepsilon \frac{dn}{d\varepsilon} \frac{d\sigma _{\gamma \gamma}}
{d\cos \theta}(E_{\gamma}, \theta, \varepsilon)
\end{equation} 
For details, see Kneiske et al. (2004).
For reference, we plot optical depths at
$z=0.1$ for the other models in Fig. 4.
Note that the simple
power-law fitting formula of Casanova et al. (2007)
overestimates optical depths in comparison with the other models 
above $\sim$TeV. 
\begin{figure}[b]
\includegraphics[width=\linewidth]{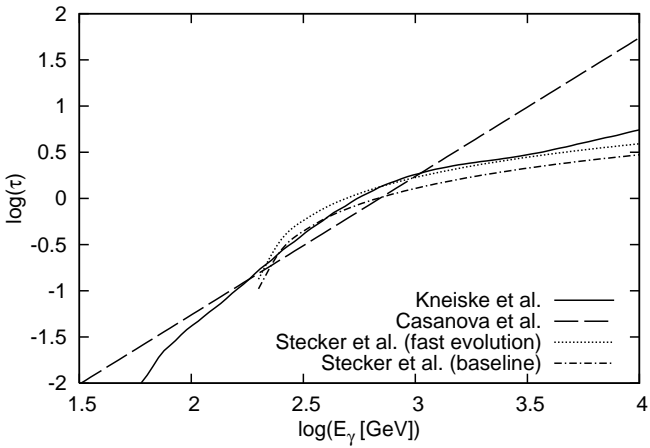}
\caption{\label{Fig4} The optical depth of high-energy gamma-rays from
a source at $z=0.1$.}
\end{figure}

\subsection{\label{subsec:e}Diffuse Gamma-Ray Background}
As noted before, if the IGM field is strong enough ($B \gtrsim 10^{-16}$ G),
it is hard to detect delayed emissions as a source
connecting with GRB prompt emissions because of large $\Delta t_B$.
Therefore, such emissions may be detected as the diffuse
gamma-ray background emission rather than delayed emissions. Now, we 
consider this extreme case.

The observed diffuse gamma-ray emission was found to be a
power-law in energy and is highly isotropic on the sky \cite{Sre1}, 
but it may not be consistent with a simple power-law and 
The origin of the diffuse gamma-ray background from extragalactic 
sources is also an open question. Blazar is one of the most discussed and 
promising candidates. Other sources such as fossil radiation from accelerated
cosmic rays during the structure formation might give a significant 
contribution. GRB is one of the brightest astrophysical phenomena and also
can contribute to the gamma-ray background. GRBs as sources for the $\sim$MeV
gamma-ray background were given by Hartmann et al. (2002). Casanova et 
al. (2007) considered GRBs as sources of the $\sim$GeV-TeV gamma-ray 
background. However, since GRBs are rare phenomena despite of their 
brightness, the contribution to the gamma-ray background will be
small, as is shown later.

We estimate the diffuse gamma-ray background,
independently of the IGM, as follows. The number of
GRBs is written as,
\begin{equation}
\dot{N}_{\mr{GRB}} = \int _{z_{\mr{min}}}^{z_{\mr{max}}} dz 
\frac{\rho _{\mr{GRB}}(z)}{1+z}
\frac{dV}{dz},
\end{equation}
where the volume factor
\begin{eqnarray}
\frac{dV}{dz} &=& \frac{c}{H_{0}}\frac{4 \pi d_{L}^2}{{(1+z)}^{2}
\sqrt{\Omega _{m}{(1+z)}^{3}+
\Omega _{\Lambda}}},
\end{eqnarray}
($d_{L}$ is the luminosity distance) and the GRB rate
\begin{eqnarray}
\rho _{\mr{GRB}}(z) &=& \rho _0 \frac{23 e^{3.4z}}{22+e^{3.4z}} 
\frac{\sqrt{\Omega _{m}{(1+z)}^{3}+
\Omega_{\Lambda}}}{{(1+z)}^{3/2}}.
\end{eqnarray}
Here, we have used the SF2 model of Porciani \& Madau (2001) for
the GRB rate with $\rho _0=1 \,
\mr{Gpc}^{-3}\mr{yr}^{-1}$, assuming that the GRB rate traces the star
formation rate in a global sense. Guetta et al. (2004) obtained such a value 
of the GRB rate and Liang et al. (2007) also reported a similar
value. Even though the actual GRB rate may
not be a good tracer of the star formation rate \cite{Gue3,Le1}, our
conclusion about the diffuse background would not be changed so much 
because the main contribution to the background 
comes from bursts that occur at $z \sim (1-2)$, the number of which is
observationally determined.
The diffuse gamma-ray background due to GRBs is estimated by,
\begin{equation}
\frac{dF_{\gamma}}{dE_{\gamma}} = \int _{z_{\mr{min}}}^{z_{\mr{max}}}
\, dz \left( \frac{dN_{\gamma}}{dE_{\gamma}dA} \right) \frac{d
{\dot{N}}_{\mr{GRB}}}{dz} \label{gammabg}
\end{equation}
where  $dN_{\gamma}/dE_{\gamma}dA$ is the observed gamma-ray fluence
from each burst, which is defined by,
\begin{equation}
\frac{dN_{\gamma}}{dE_{\gamma}dA} \equiv \frac{1}{4 \pi d_p^2} 
\frac{dN_{\gamma}}{dE_{\gamma}},
\end{equation} 
where $dN_{\gamma}/dE_{\gamma}$ is the photon number spectrum (where the
source redshift is taken into account) and $d_p$ is the proper
distance to a source.   
We set $z_{\mr{min}}=0$ and $z_{\mr{max}}=5$.

We have to note that TeV emissions from GRBs in the internal shock
model can be expected only in the
limited cases. In the context of the internal shock model, 
a sufficiently large Lorentz factor and/or large collision radius 
are required.  
Although a fraction of such GRBs that can emit TeV gamma-rays
is unknown, maybe only a fraction of GRBs are TeV emitters. 
Hence, the contribution of GRBs to the diffuse 
gamma-ray background would give an upper limit, since it is evaluated
under the assumption that all GRBs have spectra extended to TeV energies,.

\section{\label{sec:2}Results}
\subsection{\label{subsec:f}Delayed Gamma-Ray Spectra}
In Figs. 5 and 6, we show total fluences of prompt and delayed gamma-rays 
numerically calculated for various redshifts.
In model A ($E_{\gamma}^{\mr{max}}=1$ TeV), the maximum energy
of secondary pairs is at most $\sim 500$ GeV in the local rest frame.
Hence, the typical energy of up-scattered CMB photons (hereafter, USMB photons)
is $\sim 1$ GeV, above which USIB photons can make
significant contributions to resulting delayed spectra. Such
USIB photons form a relatively flat ``slope'' in spectral 
shape in the $\sim (10-100)$ GeV range for $z \lesssim
3$. For $z \gtrsim 3$, such high-energy slope signature becomes 
difficult to be seen, because secondary photons are absorbed again.

The above picture is changed, if $E_{\gamma}^{\mr{max}}$ is
beyond 1 TeV.
For model B ($E_{\gamma}^{\mr{max}}=10$ TeV), the maximum energy
of secondary pairs is $\sim 5$ TeV in the local rest frame.
Therefore, the energy of USMB photons
can reach $\sim 100$ GeV. Similarly to the case of
$E_{\gamma}^{\mr{max}}=1$ TeV, the contributions from USIB
photons are important above $\sim 100$ GeV.
However, such high-energy photons may not reach the Earth, 
because of duplicated absorption.
The optical depth against high-energy photons above $\sim 200$ GeV 
exceeds unity for $z \sim 0.4$. Hence, the effect of USIB photons will be
buried unless GRBs occur at sufficiently low redshifts. 
\begin{figure}[bt]
\includegraphics[width=\linewidth]{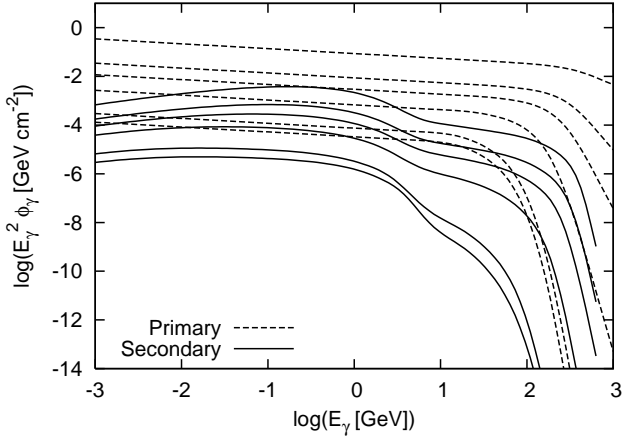}
\caption{\label{Fig5} The overall fluences of primary and secondary
gamma-rays for model A ($E_{\gamma}^{\mr{max}}=1$ TeV). 
Redshifts (from top to bottom) are $z=0.1$,
$z=0.3$, $z=0.5$, $z=1$, $z=3$ and $z=5$.}
\end{figure}
\begin{figure}[bt]
\includegraphics[width=\linewidth]{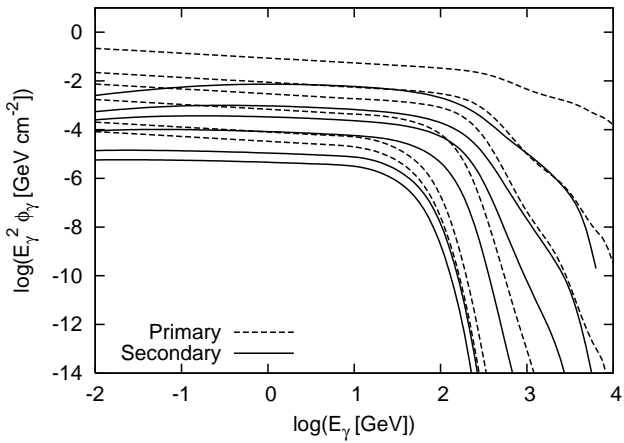}
\caption{\label{Fig6} The same as Fig. 5, but for model B 
($E_{\gamma}^{\mr{max}}=10$ TeV).}
\end{figure}

Although we have discussed CIB effects based on fluence 
$E_\gamma^2 \phi_\gamma$ in Figs. 5 and 6, flux ($F_\gamma$)
of delayed photons rather than fluence 
($E_\gamma \phi_\gamma=\int dt F_\gamma$)
is often used in discussing detectability of such photons. Hence,
in Figs. 7 and 8, we show fluxes that are 
numerically obtained for a source at $z=0.1$.
For reference, approximate results obtained by using
Eq. (\ref{delay}) are also plotted.
The delayed time scale is evaluated using Eqs. (\ref{ang})-(\ref{mag}).
In both Figs. 7 and 8, the small difference between the numerical 
results and the approximate ones using Eq. (\ref{delay}) is seen
around the peak of the bump formed by USMB photons. This
difference will be caused by cooled electrons, as explained below. 
The approximation using Eq. (\ref{delay}) means that both pair 
creation and IC scattering are treated as entirely local
processes. But high-energy electrons will be produced after passing 
$\sim \lambda_{\gamma \gamma}$ and IC scattering will occur after passing $\sim
\lambda_{\mr{IC}}= c \hat{t}_{\mr{IC}}$. Thus, these propagating
electrons suffer from IC losses as well as adiabatic losses and
effective electron distribution would deviate from the expression in 
Eq. (\ref{dist}). Photon emissivity will be also changed correspondingly, and 
the difference between two methods appears. For example, the peak
of the USMB bump becomes more ambiguous due to such
losses. Nevertheless, the approach to use Eq. (\ref{delay}) will 
usually work as a reasonable approximation.  

As is shown in Fig. 7 ($E_{\gamma}^{\mr{max}}=1$ TeV), delayed emissions in
$\sim (10-100)$ GeV due to USIB photons are prominent.
In this case, the treatment neglecting the USIB effect is
not good and leads to underestimation of the delayed gamma-ray
flux. Even for $E_{\gamma}^{\mr{max}}=10$ TeV (see Fig. 8), the USIB 
effect is still remarkable. In fact, the contribution from USIB
photons is dominant above $\sim 500$ GeV. However, the USIB effect is
smaller than the case of $E_{\gamma}^{\mr{max}}=1$ TeV, since a
fraction of delayed secondary photons is absorbed again by
CIB photons. 

USMB photons form the ``bump'' shape around the
$E_{\gamma} \sim$ a few $\times {\gamma_{e}}^2
\varepsilon_{\mr{CMB}}/(1+z)$, resembling the Planck
distribution. Unless duplicated absorption of secondary photons 
is significant, the ratio of the fluence due to USIB 
photons to that due to USMB photons should reflect the ratio of the 
CIB energy density to the CMB energy density.
This statement is basically demonstrated in Fig. 9, from which we can see that
$E_\gamma^2 \phi_\gamma$ at 100 GeV is about 0.01 times lower than 
that at 1 GeV for the normal CIB strength, when the burst occurs at
$z=0.1$. However, for the sufficiently high $z$ bursts, the duplicated 
absorption of secondary photons becomes significant, where the fluence
ratio will largely deviate from the ratio of the CIB intensity to CMB 
one. This implies that we need delayed gamma-ray spectra
over wide energy ranges in order to see the USIB effect  
on the delayed spectra, since the cutoff of delayed secondary emission 
depends on the CIB strength and the distance to the burst. Note that 
the number of CMB photons increases at high redshifts, while 
that of CIB photons does not change monotonically.
Therefore, USMB photons are more prominent for bursts at high redshifts.
\begin{figure}[tb]
\includegraphics[width=\linewidth]{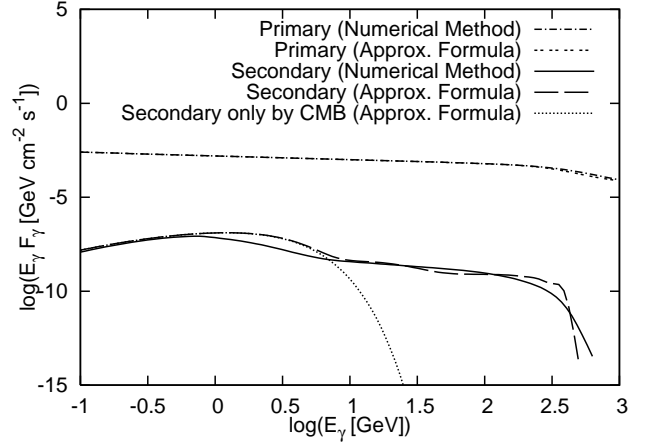}
\caption{\label{Fig7} Flux for model A ($E_{\gamma}^{\mr{max}}=1$ TeV)
obtained numerically and obtained using Eq. (\ref{delay}).
The redshift of a source is set to $z=0.1$.
The primary flux is evaluated for the intrinsic GRB duration
$T^{\prime}=50$ s (i.e. time-averaged flux over the duration is
assumed). On the other hand, the delayed secondary flux is 
evaluated at the observation time $t={10}^{4}$ s (i.e. the flux which 
is essentially time-averaged over $t={10}^{4}$ s is assumed).}
\end{figure}
\begin{figure}[tb]
\includegraphics[width=\linewidth]{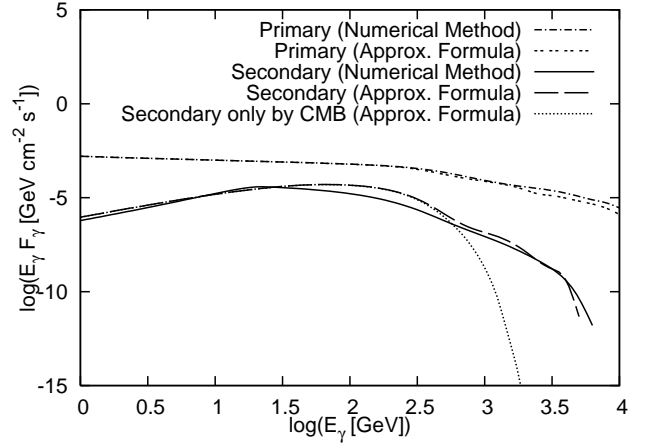}
\caption{\label{Fig8} The same as Fig. 8, but for model B
and the observation time $t={10}^{2}$ s.}
\end{figure} 
\begin{figure}[tb]
\includegraphics[width=\linewidth]{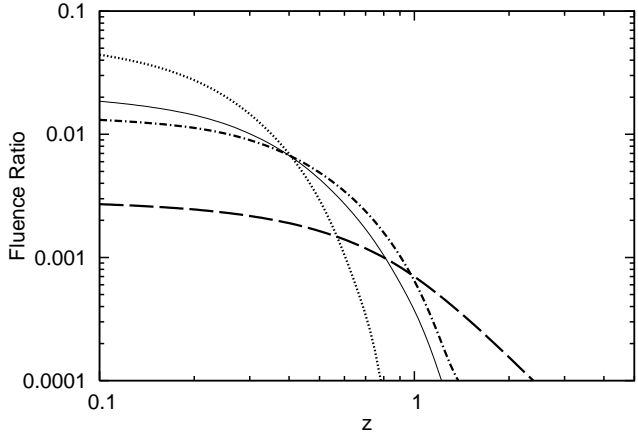}
\caption{\label{Fig9} The ratio of $E_\gamma^2 \phi_\gamma$ at 100 GeV to that at
1 GeV. The thick dot-dashed line shows the energy fluence ratios for model A and the 
thin solid line shows those for model C in the normal CIB case, where the best-fit
model given by Kneiske et al. (2004) is used.
The dashed and dotted lines show the ratios for model A in
the stronger CIB case (5 times as strong as the normal case)
and weaker CIB case (0.2 times), respectively.}
\end{figure} 

In model C, the amount of prompt TeV photons is much less than that in model A,
and the spectrum of primary prompt emission is not expressed as a
simple power-law. Nevertheless, the ratio of $E_\gamma^2 \phi_\gamma$
in model C is similar to that in model A.
Thus, the 100 GeV-1 GeV ratio for low redshift bursts
is a good indicator of the CIB strength
irrespective of the spectral shape of prompt emissions,
unless the amount of photons above TeV is considerable.

In Figs. 10 and 11, we show time-integrated fluxes, i.e., fluences at
a given time for the model A from sources at $z=0.1$ and $z=1$, respectively. 
In both figures, we change the overall CIB intensity by
a factor 5 and 0.2 in order to demonstrate effects of the CIB intensity.
As we change the strength of the CIB field, the delayed secondary 
fluence also changes correspondingly, which is also shown in Fig. 9.
For $z=0.1$, the cosmic space is essentially optically thin for pair 
creation against CIB photons, so that secondary gamma-rays are not 
completely absorbed. The USIB effect is outstanding above $\sim (10-100)$ GeV
range. Even around the bump, where USMB photons are
dominant, the fluence changes according to the amount of
absorbed primary photons, which reflects the CIB intensity. We can
also see the influence of duplicated absorption in the high-energy
range, although it is small for the case with $E_{\gamma}^{\mr{max}}=1$ TeV
and $z=0.1$.   

For $z=1$, delayed secondary spectra become more complicated. Above
$\sim 70$ GeV, gamma-rays cannot reach the Earth without
attenuation. Hence, delayed secondary gamma-rays above this energy are
absorbed again and regenerated. If we change the strength of the CIB field
(as represented in the dotted or dashed line in Figs. 10 and 11), the cutoff
energy, at which the optical depth becomes $\tau_{\gamma \gamma}^{\mr{bkg}}=1$,
also changes. 
Although the height of the USMB bump is also affected by the CIB intensity, 
this influence will be saturated when primary gamma-rays are
completely attenuated, like in the cases shown in Fig. 11.
\begin{figure}[tb]
\includegraphics[width=\linewidth]{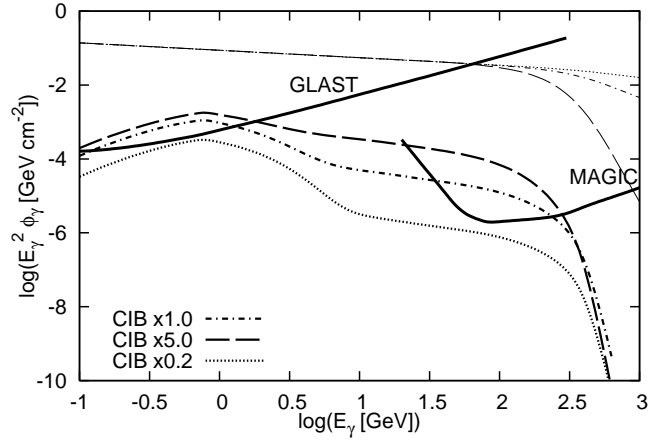}
\caption{\label{Fig10} The fluences of primary and secondary
gamma-rays. Model A with the redshift $z=0.1$ is used. Fluences of prompt 
primary (upper three thin lines) and delayed secondary emission (lower three
thick lines) are time-integrated over 
$T^{\prime}=50$ s and $t={10}^{4}$ s, respectively.
The CIB strength is assumed to be normal (the best-fit model, dot-dashed),
stronger (5 times, dashed), and weaker (0.2 times, dotted). 
}
\end{figure}
\begin{figure}[tb]
\includegraphics[width=\linewidth]{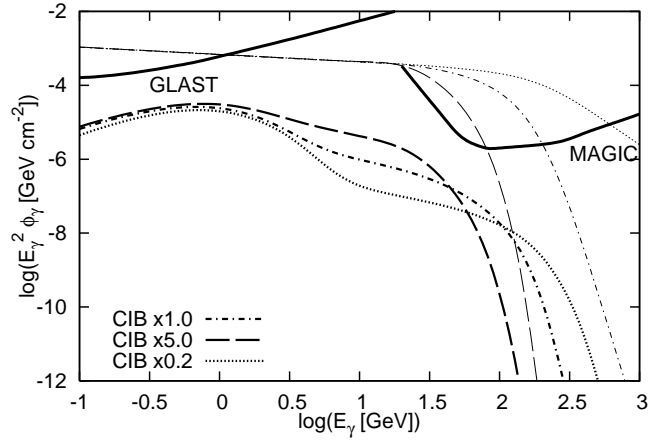}
\caption{\label{Fig11} The same as Fig. 10, but for $z=1$.}
\end{figure}     

Figs. 12 and 13 show the contrastive results among the three models. 
For low redshift bursts as
shown in Fig. 12, the high-energy slope feature is produced by the CIB for all
the models. 
For high redshift bursts, such high-energy USIB slope appears for 
model A and model C, which is shown in Fig. 13.
For model C, although a bump
signature formed by USMB photons is difficult to be seen, 
the USIB effect is still important and delayed secondary photons 
above $\sim 5$ GeV comes mainly from such USIB
photons. On the contrary, in the case of model B,
USIB photons are almost completely absorbed again. Hence, the
contribution of USIB photons is negligible and it is sufficient to consider
USMB photons only in this case. 
\begin{figure}[tb]
\includegraphics[width=\linewidth]{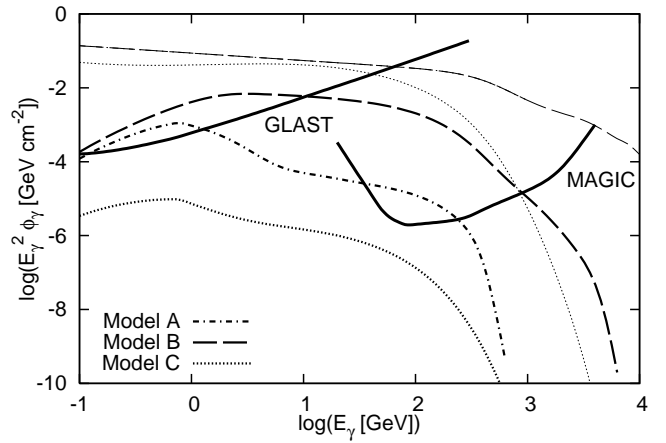}
\caption{\label{Fig12} The same as Fig. 10, but for models
A, B and C with normal CIB strength. The dot-dashed and dotted lines are 
degenerate for prompt primary spectra. }
\end{figure}
\begin{figure}[tb]
\includegraphics[width=\linewidth]{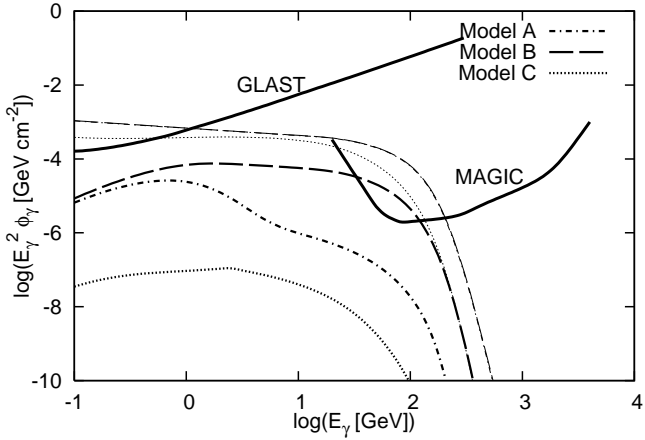}
\caption{\label{Fig13} The same as Fig. 12, but for $z=1$. The
dot-dashed and dotted lines are degenerate for prompt primary spectra.}
\end{figure}

In Figs. 10-13, we have also shown the short-time sensitivity curves of 
GLAST and MAGIC just for comparison. For GLAST, the fluence 
threshold is roughly proportional to ${(t/2.4 \times {10}^{4} \, 
\mr{s})}^{1/2}$ for the long-time sensitivity regime (exposure time 
$t \gtrsim 2.4 \times {10}^{4}$ s), and is roughly constant for the 
short-time sensitivity regime (exposure time $t \lesssim 2.4 \times 
{10}^{4}$ s). The short-time sensitivity is calculated from 
the effective area of LAT onboard GLAST (which is shown in 
``\textit{http://www-glast.slac.stanford.edu/software/IS/glast\_lat\_performance.htm}''), 
under the criterion that at least 5 photons are collected (Kamae 2006, Private
Communication). Note that the GLAST sensitivity curves shown in some
references \cite{Pet1,Raz1} are overestimated. For MAGIC, 
the short-time sensitivity is roughly estimated from the effective area 
for the zenith angle with $20$ degree by the criterion that 
at least 10 photons are collected, although the actual detectability 
requires careful analyses \cite{MAG2}. In addition, the duration time
of delayed signals can be much longer than that of prompt signals. 
For such long lasting signals, we have to take into 
account the fact that the fluence sensitivity is proportional to 
$t^{1/2}$ for the long-time sensitivity regime.  

For $10^{53}$ ergs bursts at $z=0.1$, delayed signals can be detected 
by GLAST for model A and model B. In addition, the USIB effect 
around $(10-100)$ GeV in delayed emissions could be detected by MAGIC if 
the CIB strength is strong enough. 

Even for the case of $z=1$, MAGIC and GLAST have possibilities
to detect prompt primary signals (although MAGIC has not observed
such high-energy emission up to now), if GRBs with $10^{53}$ ergs
are $\sim 10$ GeV - $10$ TeV emitters. However, even GLAST could not 
see delayed secondary components for $z \gtrsim 1$ unless bursts are
much more energetic. This result is different from 
that of Razzaque et al. (2004). This discrepancy comes from just their 
overestimation of the GLAST sensitivity. Only energetic and/or 
low redshift bursts allow GLAST to detect secondary 
delayed signals from GRBs. 

\subsection{\label{subsec:g}Diffuse Gamma-Ray Background}
In the previous section, we have considered the weak IGM field case. 
Here, let us consider the opposite case, where secondary gamma-rays
contribute to the diffuse gamma-ray background.
Fig. 14 shows the resulting diffuse background in the sense of
cumulative gamma-ray background. 
Even if the primary photons are assumed to contribute
to the diffuse background, the contribution
is much smaller than the EGRET limit.
The contribution due to delayed emissions
is much less important.
This is easily understood as
follows. If we assume the local GRB rate $\rho_{\mr{GRB}}$
(without beaming correction) $\sim 1 \,
\mr{Gpc}^{-3} \mr{yr}^{-1}$ and the released isotropic energy 
$E_{\mr{iso}} \sim {10}^{53}$ ergs,
the Hubble time $t_{\mr{H}}$ ($\sim {10}^{10}$ yr) and the
possible cosmological evolution factor on the rate leads to the
diffuse background $E_{\gamma}^2 \Phi_{\gamma}\sim (1/4\pi) c
E_{\mr{iso}} \rho_{\mr{GRB}} (z=1) t_{\mr{H}} \sim {10}^{-7} \, \mr{GeV}
\mr{cm}^{-2} \mr{s}^{-1} \mr{sr}^{-1}$, 
which is much smaller than the EGRET limit
$\sim {10}^{-6} \, \mr{GeV} \mr{cm}^{-2} \mr{s}^{-1} \mr{sr}^{-1}$ at
GeV. Hence, the contribution from GRBs to gamma-ray background is
expected to be at most $\sim 10$ \%. 
\begin{figure}[bt]
\includegraphics[width=\linewidth]{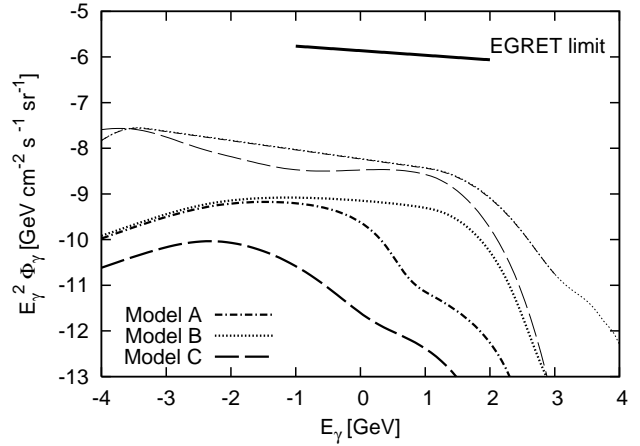}
\caption{\label{Fig14} The possible diffuse gamma-ray background from
GRBs for each model. It is assumed that the GRB rate traces the SF2
model with the local rate $\rho_{\mr{GRB}}=1 \, \mr{Gpc}^{-3}
\mr{yr}^{-1}$. Note that the contribution from delayed secondary
emission is expected to be smaller than that from prompt primary one. 
The dot-dashed and dotted lines are degenerate for prompt primary
spectra.}
\end{figure}

\section{\label{sec:3}Summary and Discussion}
In this paper, we have studied the delayed secondary emission in detail.
We have evaluated delayed GRB spectra in most detail by numerical
simulations, which enable us to treat cascade processes including 
multiple pair creation and IC scattering. We have also
calculated delayed GRB spectra using Eq. (\ref{delay}) and compared
both results. As seen in the previous
section, both methods agree with each other basically.
  

We have especially focused on effects of the CIB. CIB photons play a
role not only on absorbing high-energy gamma-rays but also on 
being up-scattered as seed photons by created high-energy pairs.
USIB photons have larger energy than
USMB photons, so that USIB photons are more subject to duplicated
absorption by CIB photons. The USIB component is more
sensitive to the CIB than the USMB component.
One of the most frequently discussed method to probe the CIB
is measuring the depletion due to the CIB in prompt GRB spectra.
However, owing to uncertainties in GRB intrinsic spectra,
the depletion is hard to be estimated. On the other hand, the 
USIB signature could be more useful to probe the CIB
almost irrespective of intrinsic GRB spectra as shown in
Fig. 9 for models A and C. While we focus on secondary emissions from
GRBs, secondary emissions from blazars are more expected
\cite{Aha1,Dai1,Fan1}. Our results can be also applied to blazars
and effects of the CIB can be importrant.
 
In this paper, we have clarified the USIB effect especially. 
As shown in Figs. 10-13, not only GLAST but MAGIC and VERITAS could 
detect such high-energy gamma-rays arising from USIB photons 
for sufficiently low redshift bursts. 
There are several characteristic features related to
USIB photons: (1) The USIB signature such as high-energy slope appears 
for bursts with the sufficiently low intrinsic high-energy cutoff
$E_{\gamma}^{\mr{max}}$ and/or sufficiently low redshift. 
(2) If (1) is satisfied (which means duplicated absorption can be 
neglected), the ratio 
of the USIB component to the USMB one basically reflects the ratio of 
the CIB energy density to the CMB one. (3) The cutoff of delayed
spectra is sensitive to the CIB, because secondary photons can also be 
absorbed by CIB photons again. Therefore, the energy range, in which 
USIB photons are prominent, becomes narrower, as the source redshift 
becomes higher. (4) As long as the effective intrinsic high-energy cutoff
$E_{\gamma}^{\mr{max}}$ is smaller than TeV, the shape of delayed secondary
spectra, especially the 100 GeV-GeV ratio,
is not so sensitive to the shape of prompt primary spectra.
For example, high-energy USIB slope can be found in both model A
and model C as seen in the previous section.      


It is important to know the highest energy of gamma-rays. These 
gamma-rays, which may arise from electron synchrotron radiation, electron SSC, 
proton synchrotron and particles generated by photomeson or photopair 
production, will suffer from absorption due to electron-positron pair 
creation. Numerical simulations in Asano \& Inoue (2007) show that
the intrinsic high-energy cutoff detemined by photon absorption is 
approximated as
\begin{equation}
E_{\gamma}^{\mr{max}} \approx 10^9 
\left( \frac{\Gamma}{100} \right)^{4}
\left( \frac{E_{\rm sh}}{10^{51} {\rm ergs}} \right)^{-0.5}
\left( \frac{\delta t}{1 {\rm s}} \right)^{1.3}
{\rm eV,} \label{max}
\end{equation}
where $\delta t$ is the variability time scale in prompt emission.
Hence, it is possible to
constrain a bulk Lorentz factor by observing the highest gamma-ray
energy. However, the highest gamma-ray energy can be seen in the prompt
spectrum only if attenuation due to the CIB is not significant. 
Because delayed secondary emission is also influenced by the highest
gamma-ray energy (for example, the typical energy of the USMB bump is
affected by the intrinsic high-energy cutoff $E_{\gamma}^{\mr{max}}$),
observations of these delayed signals 
could provide us with useful information on the source.  

Now, MAGIC has continued observations and given upper limits for
some events \cite{MAG1,MAG2}. Upper limits are also set by other
detectors such as Whipple \cite{Con1} and STACEE \cite{Jar1}. So far,
no excess event above $\sim 100$ GeV was detected, neither during
the prompt emission phase nor during the early afterglow. The upper
limits between $85$ and $1000$ GeV are derived and compatible with a
naive extension of power-law spectra. 

As we have discussed, the USIB signature in delayed spectra can be found
for GRBs with the relatively low redshift and/or relatively low
high-energy cutoff. For example, the rate of GRBs within $z=0.2-0.3$ is 
$\sim \mr{a \, few} \, \, \mr{yr}^{-1}$, so that we could see such 
delayed signals by detectors such as MAGIC in the near
future. However, the event rate would be not so large. The various
conditions such as a field of view and observational conditions of the 
detector will reduce the expected event rate
significantly. Furthermore, the real rate may be smaller because 
the number of bursts that can emit such high-energy gamma-rays may be
limited.
GLAST will see the significant number of GRBs emitting high-energy 
gamma-rays and provide us with information on high-energy spectra of prompt
emission. In addition, by using GLAST as the monitor of GRBs
that emit high-energy gamma-rays, opportunities to observe high
energy gamma-rays by ground
telescopes will also be increased. Since GLAST may see many bursts, some
of which may include ones with delayed $\sim$ GeV components. 
Furthermore, MAGIC-\Roman{ni} is now being constructed and 
VERITAS has started observations. These advances in
detectors could enable us to expect more and more chances to high-energy
gamma-ray signals from GRBs.

Here, we discuss possible complications.
One is possible influence of environments around
GRBs. We have neglected effects of the
environments around host galaxies. 
But, there might be influences from environments such as magnetic field
of the host galaxies.
The second is the existence of the IGM field. In
this paper, we have assumed the weak IGM field with 
$B \lesssim {10}^{-(18-19)}$ G to estimate delayed secondary fluxes, where the
magnetic deflection time is not a dominant time scale and the
angular-size spreading of delayed gamma-rays is sufficiently small. 
Although such a weak magnetic
field might be possible in the void region, it becomes difficult to
observe delayed signals when the IGM field is strong enough. This
is also the reason why the orphan delayed emission (which can be
expected when $\theta_B \gtrsim \theta_j$) is difficult to be detected.
Especially for $B \gtrsim {10}^{-16}$ G, delayed secondary photons
will be observed as isotropic diffuse signals that are difficult to be
observed.

The third possible complication would arise from
prolonged intrinsic high-energy emission. TeV
signals are also expected in the context of the afterglow theory.
Although the discrimination might be difficult, time-dependent 
observations could enable us to distinguish between
two signals, because the time evolution will be different between the
prompt emission and afterglow emission. Other possible late 
activities discovered by Swift would make further contamination. For
example, flares can be accompanied not only by
neutrinos and gamma-rays associated with flares themselves \cite{KM2}
but also by gamma-rays which forward shock electrons up-scatter \cite{Wan3}. 

Finally, we shall comment on high-energy emissions from low-luminosity (LL)
GRBs. The recent discovery of XRF 060218 \cite{Cam1} implies that
there may be a different population from usual cosmological
high-luminosity (HL) GRBs. These LL GRBs are more frequent than usual HL GRBs.
If true, the high-energy neutrino background from LL GRBs can be
comparable with that from HL GRBs \cite{KM3,Gup1}. 
Similarly, we can expect the gamma-ray
background from LL GRBs is comparable with that from HL GRBs. This
is also pointed out by Casanova et al. (2007) and Dermer (2006). However, 
the diffuse gamma-ray background from GRBs is much smaller than the
EGRET bound and the diffuse component will be very difficult to be detected. 
Furthermore, it might be difficult to emit TeV photons from LL GRBs, unless
they have large Lorentz factors.
Although we cannot deny
the possibility for LL GRBs to emit TeV photons so far, we do not consider
such cases in this paper.

In summary, we have done the most detailed numerical 
calculations, and justified the frequently used approximate
approach. We have especially studied CIB effects on delayed gamma-ray
spectra from GRBs in detail. CIB photons are important because
primary TeV gamma-rays are absorbed by them. In addition, we have
emphasized that not only CMB photons but also CIB photons will 
be up-scattered by produced electron-positron pairs, and enable
delayed secondary spectra to extend to higher energies. Although this 
USIB effect has not been emphasized in previous studies, it is
important since we could obtain additional information on the CIB. 
     

\acknowledgments
We thank T. Kneiske for giving us advice to use their CIB spectral model.
K.M. thank T. Totani. and K. Ioka for comments on effects of the
CIB. K.M. and S.N. also thank J. Granot, T. Kamae and M. Teshima for helpful
comments. K.M. and K.A. thank S. Inoue for helpful discussion. We are
grateful to Z.G. Dai and Y.Z. Fan for useful advices. We would like to 
appreciate for the anonymous referee for giving us profitable suggestions. 

This work is supported by the Grant-in-Aid for the 21st Century COE
"Center for Diversity and Universality in Physics" from the Ministry
of Education, Culture, Sports, Science and Technology (MEXT) of Japan.
The work of K.M. is supported by a Grant-in-Aid for JSPS Fellows.

\clearpage






\begin{thebibliography}{}
\bibitem[Aharonian et al. 1994]{Aha1}Aharonian, F.A. et al. 1994, ApJ,
423, L5
\bibitem[Aharonian et al. 2006]{Aha2}Aharonian, F.A. et al. 2006,
Nature, 440, 1018
\bibitem[Amenomori et al. 2001]{Ame1}Amenomori, M. et al. 2001,
AIPC. Proc., 599, 493
\bibitem[Ando 2004]{And1}Ando, S. 2004, MNRAS, 354, 414
\bibitem[Asano 2005]{Asa1}Asano K. 2005, ApJ, 623, 67
\bibitem[Asano \& Inoue 2007]{Asa2}Asano, K., and Inoue, S. 2007,
arXiv:0705.2910
\bibitem[Asano \& Nagataki 2006]{Asa3}Asano, K., and Nagataki, S. 2006,
ApJ, 640, L9
\bibitem[Asano \& Takahara 2003]{Asa4}Asano, K., and Takahara,
F. 2003, PASJ, 55, 433
\bibitem[Atkins et al. 2000]{Atk1}Atkins, R. 2000, ApJ, 533, L119
\bibitem[Blumenthal \& Gould 1970]{Blu1}Blumenthal, G.R., and Gould,
R.J. 1970, Rev. Mod. Phys., 42, 237
\bibitem[Bruzual \& Charlot 1993]{bru93}Bruzual, A. G., and Charlot, S. 1993,
ApJ, 405, 538
\bibitem[Campana et al. 2006]{Cam1}Campana, et al. 2006, Nature, 442, 1008 
\bibitem[Casanova et al. 2007]{Cas1}Casanova, S., Dingus,
B., and Zhang, B. 2007, ApJ, 656, 306
\bibitem[Cheng \& Cheng 1996]{Che1}Cheng, L.X., and Cheng, K.S. 1996,
ApJ, 459, L79 
\bibitem[Connaughton et al. 1997]{Con1}Connaughton, V. et al. 1997,
ApJ, 479, 859
\bibitem[Dai et al. 2002]{Dai1}Dai, Z. G., Zhang, B., Gou, L. J., 
Meszaros, P., and Waxman, E. 2002, ApJ, 580, L7
\bibitem[Dai \& Lu 2002]{Dai2}Dai, Z.G., and Lu, T. 2002, ApJ, 580, 1013
\bibitem[Dermer 2006]{Der1}Dermer, C.D. 2006, astro-ph/0610195
\bibitem[Dermer and Atoyan 2003]{der03}
Dermer, C. D., \& Atoyan, A. 2003, \prl, 91, 071102
\bibitem[Dermer \& Atoyan 2004]{Der2}Dermer, C.D., and Atoyan,
A. 2004, A\&A, 418, L5
\bibitem[Dermer et al. 2000]{Der3}Dermer, C.D., Chiang, J., and
Mitman, K. 2000, ApJ, 537, 785
\bibitem[Derishev et al. 2002]{Der4}Derishev, E. V., Kocharovsky,
V. V., and Kocharovsky, Vl. V. 2001, A\&A, 372, 1071
\bibitem[Enomoto et al. 2002]{Eno1}Enomoto, R. et al. 2002,
Astropart. Phys. 16, 235
\bibitem[Fan et al. 2004]{Fan1}Fan, Y. Z., Dai, Z. G., Wei,
D. M. 2004, A\&A, 415, 483
\bibitem[Guetta et al. 2001]{gue01}
Guetta, D., Spada, M., \& Waxman, E. 2001, \apj, 559, 101
\bibitem[Guetta \& Granot 2003]{Gue1}Guetta, D., and Granot, J. 2003,
ApJ, 585, 885
\bibitem[Guetta et al. 2004a]{gue04}
Guetta, D., Hooper, D., Alvarez-Mu\~niz, J., Halzen, F.,
\& Reuveni, E. 2004, Astropart. Phys., 20, 429
\bibitem[Guetta et al. 2004b]{Gue2}Guetta, D., Perna, R., Stella, L.,
and Vietri, M. 2004, ApJ, 615, L73 
\bibitem[Guetta \& Piran 2007]{Gue3}Guetta, D., and Piran, T. 2007, 
JCAP, 07, 003
\bibitem[Gupta \& Zhang 2007]{Gup1}Gupta, N., and Zhang, B. 2007,
Astropart. Phys., in press
\bibitem[Hartmann et al. 2002]{Har1}Hartmann, D.H. et al. 2002,
AIPC. Proc., 662, 477G
\bibitem[Hauser \& Dwek 2001]{Hau1}Hauser, M.G., and Dwek, E. 2001,
ARA\&A, 39, 249 
\bibitem[Hinton 2004]{Hin1}Hinton, J.A. 2004, New Astron. Rev., 48, 331 
\bibitem[Holder et al. 2006]{Hol1}Holder, J. et al. 2006, astro-ph/0611598 
\bibitem[Hurley et al. 1994]{Hur1}Hurley, K. et al., 1994, Nature, 372, 652
\bibitem[Jarvis et al. 2005]{Jar1}Jarvis, B. et al. 2005, Proc. of the 29th
International Cosmic Ray Conference, 4, 455
\bibitem[Kamae 2006, Private Communication]{Kam1}Kamae, T. 2006, Private Communication
\bibitem[Kashlinsky 2006]{Kas1}Kashlinsky, A. 2006, New Astron. Rev., 50, 208
\bibitem[Kneiske et al. 2002]{Kne1}Kneiske, T.M., Mannheim, K., and
Hartmann, D.H. 2002, A\&A, 386, 1
\bibitem[Kneiske et al. 2004]{Kne2}Kneiske, T.M. et al. 2004, A\&A,
413, 807
\bibitem[Le \& Dermer 2007]{Le1}Le, T., and Dermer, C.D. 2007, ApJ,
661, 394
\bibitem[Liang et al. 2007]{Lia1}Liang, E. et al. 2007, 662, 1111
\bibitem[Lithwick \& Sari 2001]{Lit1}Lithwick, Y., and Sari, R. 2001, ApJ, 555, 540
\bibitem[MAGIC Collaboration 2006a]{MAG1}MAGIC Collaboration, 2006,
ApJ, 641 L9
\bibitem[MAGIC Collaboration 2006b]{MAG2}MAGIC Collaboration, 2006, astro-ph/0612548 
\bibitem[M\'esz\'aros 2006]{Mes3}M\'esz\'aros, P. 2006, Rep. Prog. Phys., 69, 2259
\bibitem[M\'esz\'aros \& Rees 1994]{Mes1}M\'esz\'aros, P., Rees, M. J. 1994, MNRAS, 269, L41
\bibitem[M\'esz\'aros et al. 1994]{Mes2}M\'esz\'aros, P., Rees, M. J., 
Papathanassiou, H. 1994, ApJ, 432, 181
\bibitem[Milagro Collaboration 2007]{Mil1}Milagro Collaboration 2007,
arXiv:0705.1554
\bibitem[Mirzoyan et al. 2005]{Mir1}Mirzoyan et al. 2005, Proc. of the 29th
International Cosmic Ray Conference, 4, 23
\bibitem[Murase \& Nagataki 2006a]{KM1}Murase, K., and Nagataki, S. 2006a, Phys. Rev. D, 73, 063002
\bibitem[Murase \& Nagataki 2006b]{KM2}------. 2006b,
Phys. Rev. Lett., 97, 051101
\bibitem[Murase et al. 2006]{KM3}Murase, K., Ioka, K., Nagataki, S.,
and Nakamura, T. 2006, ApJ, 651, L5 
\bibitem[Papathanassiou \& M\'esz\'aros 1996]{Pap1}Papathanassiou, H.,
and Meszaros, P., ApJ, 471, L91
\bibitem[Peer \& Waxman 2004]{Pee1}Peer, A., and Waxman, E. 2004, ApJ, 613, 448
\bibitem[Petry et al. 1999]{Pet1}Petry, D. et al. 1999, A\&A, 138, 601
\bibitem[Poirier et al. 2003]{Poi1}Poirier, J. at al. 2003, Phys. Rev. D, 67
\bibitem[Porciani \& Madau 2001]{Por1}Porciani, C., and Madau,
P. 2001, ApJ, 548, 522
\bibitem[Plaga 1995]{Pla1}Plaga, R. 1995, Nature, 374, 30
\bibitem[Razzaque et al. 2004]{Raz1}Razzaque, S., M\'esz\'aros, P.,
and Zhang, B. 2004, ApJ, 613, 1072
\bibitem[Sreekumar et al. 1998]{Sre1}Sreekumar, P. et al. 1998, ApJ,
494, 523 
\bibitem[Stecker et al. 2006]{Ste2}Stecker, F.W., de Jager, O.C., and
Salamon, M.H. 1992, ApJ, 390, L49 
\bibitem[Stecker et al. 2006]{Ste3}Stecker, F.W., Malkan, M.A., and
Scully, S.T. 2006, ApJ, 648, 774
\bibitem[Sommer et al. 1994]{Som1}Sommer, M. 1994, ApJ, 422, L63 
\bibitem[Strong et al. 2004]{Sto1}Strong, A.W., Moskalenko, I.V., and
Reimer, 0. 2004, ApJ, 648, 774
\bibitem[Totani 1998]{Tot1}Totani, T. 1998, ApJ, 509, L81
\bibitem[Totani \& Takeuchi 2002]{Tot2}Totani, T., and Takeuchi,
T.T. 2002, ApJ, 570, 470 
\bibitem[Wang et al. 2001]{Wan1}Wang, X.Y., Dai, Z.G., and Lu,
T. 2001, ApJ, 556, 1010
\bibitem[Wang et al. 2004]{Wan2}Wang, X.Y., Cheng, K.S., Dai, Z.G.,
and Lu, T. 2004, ApJ, 604, 306
\bibitem[Wang et al. 2006]{Wan3}Wang, X.Y., Li, Z., and M\'esz\'aros,
P. 2006, ApJ, 641, L89
\bibitem[Waxman \& Bahcall 1997]{Wax1}Waxman, E. and Bahcall, J. 1997,
Phys. Rev. Lett., 78, 2292
\bibitem[Zhang 2007]{Zha1}Zhang, B. 2007, ChJAA, 7, 1
\bibitem[Zhang \& M\'esz\'aros 2001]{Zha2}Zhang, B., and  M\'esz\'aros,
P. 2001, ApJ, 559, 110
\end{thebibliography}
\end{document}